# Supermassive Black Holes at the Center of Galaxies

Christopher J. Greenwood

**Abstract**. This was my final paper for the AST 308 Galaxies class at Michigan State University. Using many sources I was able to compile a moderate amount of information concerning the evidence for, and the formation of Supermassive Black Holes.

## I. Introduction

There are many aspects of astronomy that are still unknown to us to this day. Along with the nature of dark energy and dark matter, there are still some things yet to learn about black holes. We do, however, know many things about black holes including how they may form, where they are located, and how massive they can be. Today it is thought that at least one supermassive black hole lies at the center of almost every galaxy, this is further supported by observational evidence, which will be explained in more detail in the first part of this paper. Studying this phenomenon is essential when trying to understand the internal structure of galaxies, and how they evolve over time which will be covered in the second part of this paper.

Black holes form when a star about five times as massive as our own sun runs out of fuel and ceases to sustain a nuclear fusion reaction. The enormous mass of the star cannot support itself and collapses under its own gravitational field. This causes a supernova that expels mass and radiation out into space. The remnant of this great explosion is a singularity, a point in space time infinitely massive in an infinitely small volume, which causes an infinite curvature in space-time. This occurrence is also known as a "black hole", a stellar body so massive not even electromagnetic radiation such as light can escape its gravitational field once it has crossed the event horizon (Jones 150), appropriately earning it the name "black hole". Comparatively, a supermassive black hole is similar to an ordinary black hole but with a mass of roughly one million times the mass of our own sun. With such high masses, supermassive black holes are thought to only occur at the center of galaxies.

## II. Evidence and Methods

There are many methods to use when looking for a supermassive black hole, one very accurate way is to use x-rays emitted from the region though to contain a supermassive black hole. By analyzing the x-rays emitted from the immediate center of an active galactic nucleus, astronomers can determine the mass that is concentrated within a certain radius. Usually this mass is far too large to be an ordinary star ($\approx 10^8$ $M_o$), since a star this massive would have immense luminosity and be easy to identify as such, therefore we must probe deeper to find the source. Stefanie Komossa explains the precision of using x-ray signatures to find black holes in her paper "X-ray evidence for Supermassive Black Holes at the centers of Nearby, Non-Active Galaxies":

> In active galaxies, excellent evidence for the presence of SMBHs is provided by the detection of luminous hard power-law like X-ray emission, rapid variability, and the discovery of evidence for relativistic effects in the iron-K line profile. X-ray observations currently provide the most powerful way to explore the black hole region of AGN (Komossa 2).

In the nucleus of active galaxies such as quasars, radio galaxies, and blazars, there is a lot of x-rays emitting from the accretion disk spiraling around the supermassive black hole at the center. The accretion disk is made up of gas and dust particles from many different sources inside and outside of the galaxy (Jones 151). This debris gets caught in the immense gravitational field of the black hole and begins to orbit. As the particles inside the accretion disk rotate faster and faster nearing the boundary of the black hole, collisions become more frequent and energetic. This causes energy that was once rotational velocity to be lost to friction. The energy is converted to heat and light which then causes the superheated particles to try and expand outward. The immediate outward direction is blocked by the strong pressure of the spiraling accretion disk, so it discharges the only direction it can, perpendicular to the accretion disk as high energy jets and radio lobes (Fig. 1).

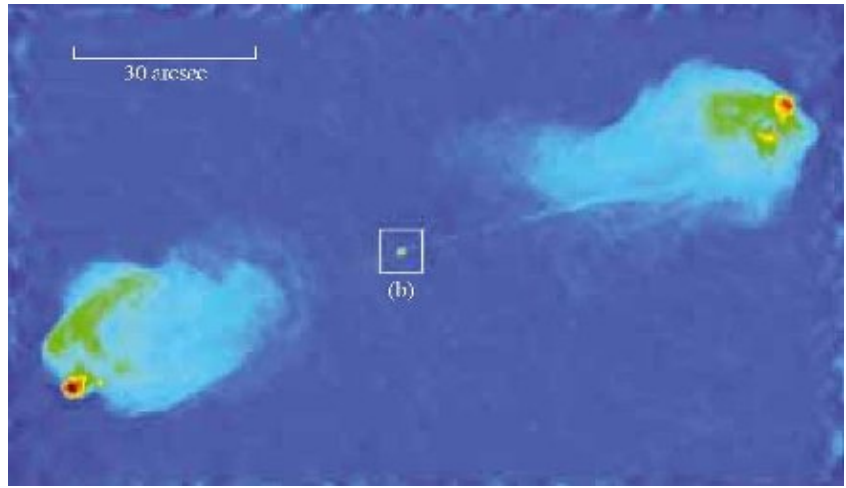

**Figure 1: This image of Cygnus A shows the radio lobes and jets caused by interactions between the accretion disk and the central supermassive black hole (Jones 142).**

As the particles near the event horizon of the black hole, significant amounts of x-rays are radiated from the source to eventually be detected by telescopes here on earth. Since black holes themselves are obviously undetectable by searching observable wavelengths, it is essential that researchers perform their search of the immediate area surrounding the black hole using observations from other methods such as x-ray telescopes and radio telescopes.

Telescopes in space such as the Hubble Telescope are extremely useful in detecting supermassive black holes from other galaxies as well as our own. The strongest evidence for supermassive black holes comes from the center of our Milky Way galaxy. Using the Hubble Space Telescope, it has been determined that objects within .2 pc of the center of Sagittarius A* (center of the Milky Way) have a rotational velocity of up to 1350 ± 40 km s$^{-1}$. Velocities close to this have also been measured on a number of other objects with similar radii (Fig. 2). Speeds such as this could only be achieved if the mass within this radius was on the order of $(2.6 \pm 2) \times 10^6$ M$_o$ (Kormendy 3).

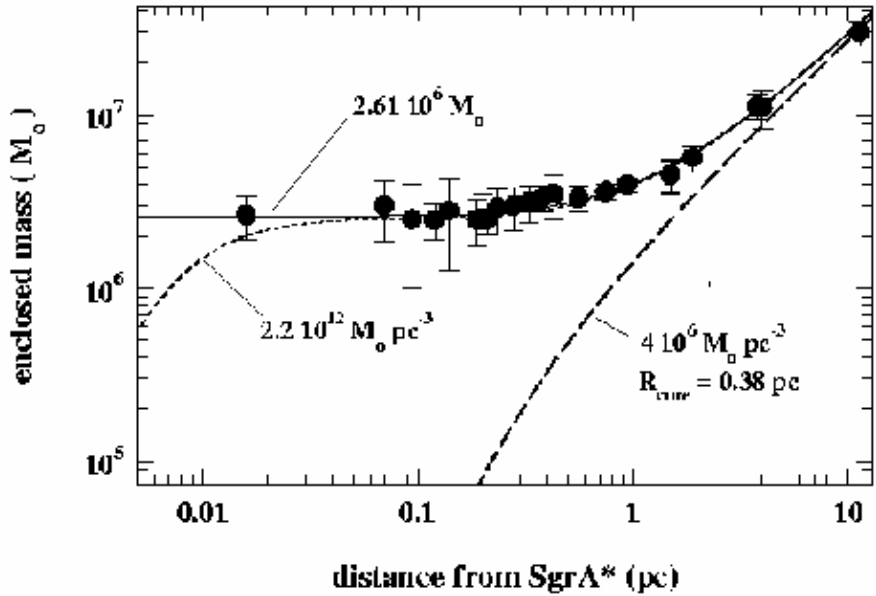

**Figure 2: This graph shows data from many stars orbiting around Sagittarius A and the enclosed mass calculated for each, they fit closely to the predicted mass of the supermassive black hole (Vaughan).**

Masses this large could not be anything other than an extremely luminous star, which would be easy to detect in the visible spectrum and would quickly exhaust its fuel and collapse into a black hole in any case. Another find is that the mass of the supermassive black hole at the center of nearby (possibly all) galaxies is proportional to the mass of the central bulge of the galaxy. This correlation has been supported by various methods such as ionized gas emission lines, ionization models, stellar proper motions, reverberation mapping, and ionized gas and emission lines (Combes 2). Another similar situation to our Milky Way is NCG 4258. This galaxy has been found to contain $4 \times 10^7$ M$_o$ within the same .2 pc as with the Milky Way (Kormendy 3).

Studying the rotational velocity and radius of objects near the center of a galaxy is one way to discover and describe supermassive black holes with relatively high accuracy, although this technique can only be used on relatively close galaxies. Another less common method relies on the idea of gravitational lensing first predicted by Einstein's theory of general relativity. The effect of a large mass on space-time causes it to warp like placing a heavy ball on a stretched out piece of rubber. And because light always follows the shortest path from one point to another, light can be bent around highly massive objects in an effect known as gravitational lensing. By determining the degree that the original image is bent around the lensing galaxy, we can then determine the mass of the galaxy's nucleus according to general relativity. This technique can be used to determine the mass of black holes at more intermediate redshifts ($0.2 \leq z \leq 1.0$) to surprising accuracy (Rusin 2). There are two different types of instances when gravitational lensing occurs depending on the mass of the black hole, either the central image is destroyed or there are two instances of the central image displayed. This technique can only be used on the latter when an additional central image is created, thus making it a far less common method for determining mass (Rusin 2). Unfortunately at this time, this method of gravitational lensing has not been used with any real data. There has yet to be discovered a pair of central images that fit the criteria that this procedure requires, but according to simulations this method works excellent and it is just a matter of time before it is put into practice. More powerful radio telescopes in the near future should provide adequate data to utilize such a technique.

Using gravitational lensing, like everything else, has a certain amount of error associated with it. David Rusin describes in his paper how systematic errors can affect the data and possible ways to overcome them:

> Despite the systematic error, we believe that the lensing technique will be useful. We are not aware of any other proposed method for the direct measurement of black hole masses in distant, inactive galaxies. … The systematic error could be eliminated by directly measuring the inner luminosity (and hence mass) profile of the lens galaxy, although this will be very difficult (Rusin et. al. 6).

Despite systematic errors the central lensed images technique sounds very promising and will hopefully be used in the very near future to help us better understand supermassive black holes and galaxy structure in general.

**III. Formation and Evolution**

One of the more speculative areas of astronomy is galactic formation and evolution. Active galaxies are all though to contain a supermassive black hole as the driving engine in the active galactic nucleus. There is also evidence that they are present in nearly all non-active galaxies also, but are much harder to detect than when they are active. One theory is that all galaxies go though an active phase in their evolution and therefore would contain a supermassive black hole even though they are not active at this stage.

Quasi-stellar objects have been observed at redshifts of $z > 6$, whose luminosities have lead astronomers to believe they house a supermassive black hole. This suggests that supermassive black holes could have been created before the universe was even a billion years old, much earlier than expected. One suggested solution to this problem is that these black holes could have been created in the early universe as the remnants of Population III stars (Combes 4). According to specific calculations and models, collapse of early stars would create many intermediate-mass black holes and binary systems.  This alone would not account for the presence of supermassive black holes, "Also, it is found that hierarchical merging can only be responsible of 10% of the total mass of present SMBH, and that gas accretion should be responsible for the rest" (Combes 5).  This shows that there are still fundamental things we have yet to discover about supermassive black holes.

There is also evidence that supermassive black holes form from galactic mergers.  The Chandra X-ray Observatory has taken images of NGC 6240, an active galaxy about 400 million light years from the Milky Way.  According to the x-ray images taken of the galactic nucleus of NGC 6240, this galaxy is the first identified to have two supermassive black holes at its center (Fig. 3).

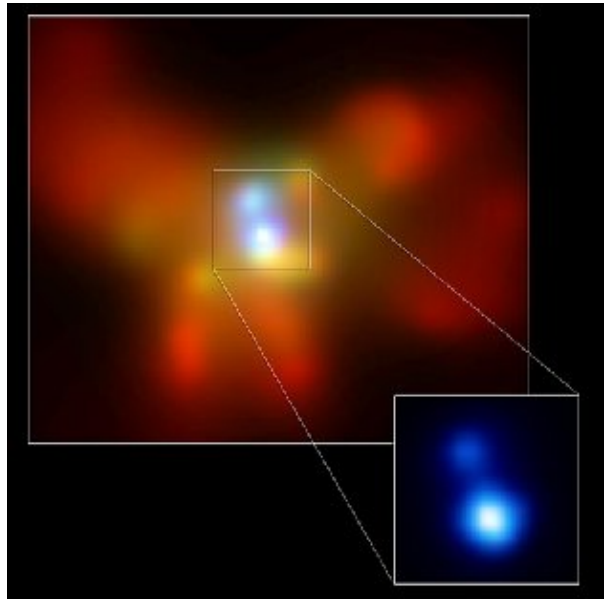

**Figure 3: X-ray image taken by Chandra of NGC 6240, first galaxy though to have two supermassive black holes at its center (Chandra NGC 6240).**

These two sources are though to be black holes because they have the necessary elements, "an excess of high-energy photons from gas swirling around a black hole, and X-rays from fluorescing iron atoms in gas near black holes" (Chandra NGC 6240).  The two supermassive black holes are 3,000 light years apart and will eventually (in a few hundred million years) collide and become an even more massive object (Chandra NGC 6240). When galaxies collide many stars are unaffected due to the fact that there is so much space in between each solar system, but when two bodies of such high mass merge together, the immense gravitational field will slingshot many smaller objects great distances. Two galaxies colliding is not an unheard of occurrence considering that our own Milky Way and our closest neighbor Andromeda will some day meet just as the two galaxies that now make up NGC 6240 once did. This observation supports the idea that supermassive black holes can form from the collision of supermassive black holes in other galaxies.

Not only can a galactic nucleus contain more than one supermassive black hole, but there may also be more black holes nearby. The Chandra X-ray Observatory has discovered an amazing number of ordinary black holes orbiting around the supermassive black hole at the center of the Milky Way. They have found evidence that this group could consist of 10,000 or more black holes within 70 light years of Sagittarius A*, making this the highest concentration of black holes in the galaxy (Chandra Swarm). They identified the black holes with x-ray images taken while monitoring the Sagittarius A* region (Fig. 4).

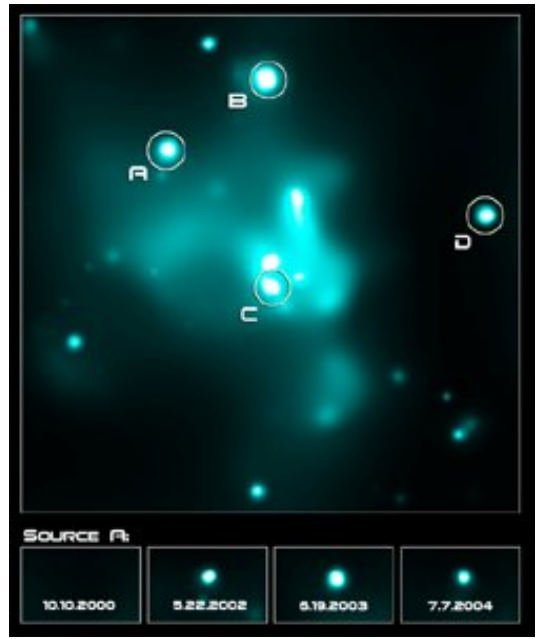

**Figure 4: The X-ray image taken at Chandra X-ray Telescope identifying a swarm of black holes orbiting around Sgr A\* (Chandra Swarm).**

The findings suggest that many black holes from the Milky Galaxy have been migrating towards the center. This action could be easily described by incorporating the gravitational effect of stars into the explanation:

> Black holes orbiting the center of the Galaxy at a distance of several light years will pull on surrounding stars, which pull back on the black holes. The net effect of this gravitational action and reaction is to decelerate the black holes, which have masses of about 10 Suns, and speed up the lower-mass surrounding stars (Chandra Swarm).

The process described above has been predicted in the past and this finding might be the evidence astronomers have been looking for. The discovery of a possible collection of black holes near Sagittarius A\* could describe the process of supermassive black hole formation. As these smaller black holes lose rotational velocity they should eventually be engulfed by Sagittarius A\*, adding to its mass. This is helpful in explaining how ordinary black holes become supermassive.

**IV. Conclusion**

There is far too much evidence in the case of supermassive black holes residing at the center of galaxies (especially active galaxies) to ignore. And although the strongest evidence is for nearby galaxies, stronger telescopes and more accurate detection methods should soon provide much more evident from galaxies at higher redshifts. This will surely lead to a clearer and deeper understanding of galaxy formation and evolution.